\documentclass[prl,twocolumn,showpacs,preprintnumbers,amsmath,amssymb]{revtex4}
\usepackage{graphicx}    
 
\begin{document} 
 
\title{Simulation of induction at low magnetic  
Prandtl number} 
 
\author{Yannick PONTY and H\'el\`ene POLITANO} 
\affiliation{CNRS, UMR 6529, Observatoire de la C\^ote d'Azur 
BP 4229, Nice Cedex 4, France}
\author{Jean-Fran\c{c}ois PINTON}
\affiliation{CNRS, UMR 5672, Laboratoire de Physique,
\'Ecole Normale Sup\'erieure, 46 all\'ee d'Italie 69007 Lyon,  France} 
 
\begin{abstract} 
We consider the induction of magnetic field in flows of electrically conducting fluid at low magnetic Prandtl number and large kinetic Reynolds number. Using the separation between the magnetic and kinetic diffusive lengthscales, we propose a new numerical approach. The coupled magnetic and fluid equations are solved using a mixed scheme, where the magnetic field fluctuations are fully resolved and the velocity fluctuations at small scale are modelled using a Large Eddy Simulation (LES) scheme. We study the response of a forced Taylor-Green flow to an externally applied field: tology of the mean induction and time fluctuations at fixed locations. The results are in remarkable agreement with existing experimental data;  a global $1/f$ behavior at long times is also evidenced.   
\end{abstract} 

 
\date{November 26, 2003} 
\maketitle

One of the strongest motivation in the study of non-linear effects in magnetohydrodynamics is that  electrically conductive flows are capable of dynamo action: the stretching of magnetic field lines by the flow velocity gradients can exceed the (Joule) diffusion. A bifurcation threshold occurs, above which the self-generation of a magnetic field takes place.  It has been validated in constrained flows  
of liquid sodium which mimic analytical models:  the Karlsruhe~\cite{karl1} and Riga experiments~\cite{riga1}.   The self-generation of a magnetic field in non-constrained homogeneous flows is still an open problem actively studied by many groups~\cite{gydroissue}.  In this research, numerical studies have long played an important role. Kinematic dynamo simulations assume a given pattern of a stationary  velocity field and study the initial linear growth rate of magnetic field perturbations. 
They have been extensively used to test  the dynamo capacity of flow geometries and  proved to be successful at determining the dynamo threshold in the Karlsruhe and Riga experiments~\cite{tilgner,galaitis}.  They have also shown that dynamo action is a possibility  in unconstrained homogeneous flows  of the von K\'arm\'an type~\cite{dudjames,marie}.  Another numerical approach  is to perform Direct Numerical Simulations (DNS) of the full governing equations: the induction equation coupled with the fluid dynamical one by the Lorentz force,  the flow being sustained by a given force (or equivalently an average geometry). They have confirmed that dynamo action 
is present in flows with differential rotations and helicity~\cite{mene, nore,nore2}.  
However, DNS are at present restricted to situations where the magnetic  
Prandtl number, $P_m = \nu/\lambda$ (where $\lambda$ is the magnetic diffusivity) 
is of order one, {\it i.e.}  to situations where the smallest scales of the magnetic and velocity fields have the same characteristic size~\cite{scheko}.  This is not the case in liquid metals, which have very small magnetic Prandtl number values {\it e.g.} $P_m \sim 10^{-6}$ for liquid Gallium and $P_m \sim 10^{-5}$ for liquid Sodium. Recall that, below the dynamo threshold, a stationary forced flow with a power input $\epsilon$ (in Watts per kg) has a viscous dissipative scale $\eta_u \sim (\nu^3/\epsilon)^{1/4}$ and a Joule diffusive scale  $\eta_B \sim (\lambda^3/\epsilon)^{1/4}$ --- hence a ratio  $\eta_u/\eta_B  \sim P_m^{3/4}$. Therefore, at low  $P_m$, the magnetic diffusive length scale  is very much larger than the velocity dissipative scale. If non-linear effects are to develop, the magnetic  Reynolds number $R_m \sim UL/\lambda$  (where $U$ and $L$ represent the characteristic velocity and scale of the flow) must be at least of order one  and thus the kinetic Reynolds number of the flow, $Re  \sim  UL/\nu  \sim  R_m / P_m$, must be very large (turbulence is fully developed). A DNS treatment of such a system is at present out of reach. 

In this paper, we present a new approach for the study of the magnetic induction in large $Re$ - low $P_m$ flows; we restrict ourselves to regimes below the dynamo threshold. In this parameter region, the magnetic field ``lives" essentially within the large and inertial hydrodynamic scales.  We thus propose to treat with a sub-grid model the velocity scales which are smaller than the magnetic diffusive length.  Schemes using hyperviscosity have previously been used~\cite{tilgner,glatzmaier}. Here, we prefer a LES approach,  which has proved very successful for the simulation of turbulent flows with large scale structures and for the modelling of energy transfers~\cite{cholletlesieur}.  In this hybrid scheme, we solve the induction equation on a  fully resolved grid and we use a LES method for the velocity field, with a cut-off scale at the end of the magnetic diffusive range.  We consider  the response of a conductive fluid  to an uniform magnetic field:  topology of the mean induced field and spatio-temporal features of the magnetic fluctuations are studied.  The chosen flow is a forced Taylor-Green vortex (TG). It shares many similarities with the experimental von K\'arm\'an swirling flows which have already been investigated in DNS near $P_m \sim {\cal O}(1)$~\cite{nore,nore2}.

In non-dimensional form, the incompressible MHD equations have two 
classical control parameters, the magnetic and kinetic Reynolds numbers, and one has to choose a forcing mechanism  that generates the desired values of  $R_m$ and $Re$.  
In order to be closer to experimental procedures, we prefer to fix the driving force  and 
the magnetic Prandtl number. Hence, the dynamical time $t_0$ is set to the magnetic 
diffusion time scale, {\it i.e.} $t_0 \lambda / {L}^2 \sim {\cal O}(1)$, where $L$ is a length scale characteristic of the system size.  Changes in magnetic diffusivity for real fluids  would change that time scale. We write the MHD equations, with constant unit density, as
\begin{eqnarray} 
\partial_t {\bf u} + {\bf u}.\nabla {\bf u} &=&   
- \nabla P + {P_m}\nabla^2 {\bf u} + {\bf F}  
+ (\nabla\times{\bf b})\times{\bf B} 
\label{eq:syst1} \\ 
\partial_t {\bf b} &=&   
\nabla \times ({\bf u} \times {\bf B}  ) 
+ \nabla^2 {\bf b} \ ,
\label{eq:syst2} \\ 
\nabla. {\bf u}&=&  0 \; \; , \; \;  \nabla.{\bf b}  =  0 \ , 
\label{eq:syst3}
\end{eqnarray}
where ${\bf u}$ is the velocity field, ${\bf B} = {\bf B}_0 + {\bf b}$  is the net magnetic field in the flow, sum of the applied and induced fields. Once the amplitude $F$ of the driving force is fixed, the (non-dimensional) $rms$ intensity of the velocity fluctuations is $u_{rms} \sim \sqrt{F}$, the Reynolds number is $Re \sim \sqrt{F}/P_m$ 
and the magnetic Reynolds number is $R_m \sim \sqrt{F}$. When the interaction parameter, ratio of the Lorentz force to the inertial forces, defined as $N \simeq  B_0^2 / u_{rms} \sim  B_0^2 / \sqrt{F} $ is small, the back reaction of the induced field on the velocity field is negligeable. The above expressions are only dimensional estimates; in practice, the characteristic flow quantities are computed as mean temporal values from the data -- cf. Table~1.
 
We use a parallelized pseudo-spectral code in a $[0-2\pi]^3$ periodic box. Time stepping is done with an exponential forward Euler-Adams-Bashford scheme. The LES model is of the Chollet-Lesieur type~\cite{cholletlesieur} in which the kinematic viscosity ${\nu}$ is replaced in spectral space by an eddy viscosity.  In Eq. (\ref{eq:syst1}) the magnetic Prandtl number is then replaced by:
\begin{equation} 
P_m(k,t) = 0.1 (1 + 5 (k/K_c)^8) \sqrt{E_v(k=K_c,t)/K_c} \ .
\label{eq:chollet-lesieur} 
\end{equation} 
Here $K_c$ is the cut-off wavenumber of the velocity field, and $E_v(k,t)$ is the one-dimensional kinetic energy spectrum. The effective  Prandtl number $P_{m_{\rm eff}}$ is obtained  as the temporal mean of $P_m(0,t)$. Note that the effective fluid viscosity $\nu_{\rm eff}$ is of the same magnitude. A consistency condition for our approach is that the magnetic field fluctuations are fully resolved when $2\pi/K_c$ is smaller than the magnetic diffusive scale  $\eta_B \sim l_0/R_m^{3/4}$, $l_0$ being the integral scale computed from the kinetic energy spectrum.  The flow is driven by the TG vortex geometry 
\begin{equation} 
{{\bf  F}_{\rm TG}(k_0)}= { 2F } \, 
\left[ \begin{array}{c} 
\sin(k_0~x) \cos(k_0~y) \cos(k_0~z) \\ 
- \cos(k_0~x) \sin(k_0~y) \cos(k_0~z)\\ 0  
\end{array} \right] 
\label{eq:Ftg}
\end{equation} 
$(k_0,k_0,k_0)$ is the wavevector that prescribes the velocity large scale (hereafter $k_0=1$). The ${\bf  F}_{\rm TG}$ and  ${\bf B_0}$ amplitudes are chosen such that 
the interaction parameter N remains smaller than $10^{-2}$.  After an initial transient ($t<10$) the flow has reached a steady state: the kinetic energy fluctuates less than 3.5\% around its mean value. All quantities are tracked up to $t_{\rm max} = 410 t_0$ --- note that $200 t_0$ is of the order of the measurement time in most Sodium experiments~\cite{lathropspec,vkspof,karlspec}. For comparison, the eddy turnover time $\tau_{NL} \sim l_0/u_{rms}$ is given in Table 1.
\begin{table}[ht] 
\begin{tabular}{|c|c|c|}\hline 
RUN  			&{\bf \#1}  $~{\bf B_0}=0.1 ~{\bf \hat {x}} $    &{\bf \#2}  
$~{\bf B_0}=0.1 ~{\bf \hat {z}}$   \\ \hline  
$TG$ 			& $Re = 9209$          	    & $Re = 9212$ \\ 
$k_0=1$ 		& $R_m = 6.65 $ 	            & $R_m = 6.68$ \\ 
$F = 3/2$                 & $R_{l_T}= 95.94$             & $R_{l_T}= 95.96 $  \\
$128^3$ grid points   & $P_{m_{\rm eff}} \sim 7.22~ 10^{-4}$ & $P_{m_{\rm eff}} = 7.26~  10^{-4} $ \\ 
$K_c = k_{\rm max} -3$	& $N= 8.23 ~ 10^{-3}$	            & $N= 8.18 ~ 10^{-3}$ \\ 
$k_{\rm max}=64$       	&$l_0= 2.338$	                    & $l_0= 2.337$ \\ 
$t_{\rm max}=410$	&$l_T = 0.024$	            & $l_T = 0.024$ \\ 
       			&$\eta_B = 0.565$	            & $\eta_B = 0.563$ \\ 
       			&$\tau_{NL} = 1.217$	    & $\tau_{NL} = 1.224$ \\ 
                	&$u_{rms}= 2.843$                & $u_{rms}= 2.858$ \\ 
       			&$b_{rms}=0.061$ 	            & $b_{rms}= 0.064$ \\  
       			&${\rm max}|{\bf u}|= 8.211$	    & ${\rm max}|{\bf u}|= 8.249$ \\ 
       			&${\rm max}|{\bf b}|= 0.160$	    & ${\rm max}|{\bf b}|= 0.180$ \\ \hline 
\end{tabular} 
\caption{Time averaged quantities: 
$u_{rms}={\langle {\bf u}^2\rangle}^{1/2}$, $b_{rms}={\langle {\bf b}^2\rangle}^{1/2}$,
flow integral scale $l_0 = 2 \pi \sum_k E_v(k)/k / \sum_k E_v(k)$, Taylor microscale $l_T \sim l_0 R_e^{-1/2}$, diffusive scale $\eta_B$ and eddy turnover time $\tau_{NL}$. 
Non-dimensional parameters: effective Prandtl number $P_{m_{\rm eff}}$,
kinetic Reynolds number $R_e=l_0 u_{rms}/\nu_{\rm eff}$ (see text),
and magnetic Reynolds number $R_m=P_{m_{\rm eff}}R_e$, Taylo-based Reynolds number $R_{l_T} \sim R_e^{1/2}$, interaction parameter $N=R_m B_0^2 / u_{rms}^2$. }
\end{table}

Figure~1 shows the power spectra of the velocity and magnetic field fluctuations with  ${\bf B_0}$ applied along the ${\bf \hat{x}}$-axis (a direction perpendicular to the rotation axis of the counter-rotating eddies of the TG cells). The kinetic energy spectrum exhibits the $k^{-5/3}$ Kolmogorov scaling law maintained throughout the range by the LES scheme. The peak at low wavenumber is due the large scale TG forcing, also visible on the magnetic field spectrum.  The magnetic inertial range is well fitted by a $k^{-11/3}$ power law in agreement with a Kolmogorov phenomenology~\cite{moffatt,odier}.  
The magnetic diffusive scale is reached within the computational box.  The main goal of our numerical strategy is thus achieved:  the magnetic fluctuations are fully resolved in a range of scales at which the velocity field follows the Kolomogorov self-similar structure of turbulence. Hence, we get the possibility to study magnetic induction in a fully developped turbulent flow at low magnetic Prandtl number.
\begin{figure} [h]
\includegraphics[width=7cm]{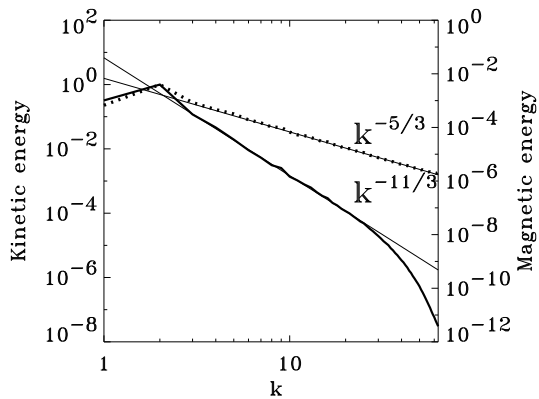} 
\caption{Magnetic (solid line) and kinetic (dash line) energy spectra computed at
$t=210$ for RUN 1 with ${\bf B_0}=0.1~{\bf {\hat{x}}}$.}
\end{figure} \\

Figure~2 displays isosurfaces of the local induced magnetic energy ${\langle E_{b}({\bf x},t)\rangle}_T$ averaged in the time interval $T=[10-410]$, shown at $80 \%$ of its maximum value. For comparison, we  also plot isosurfaces of the induced magnetic energy, ${\langle E_{b,lin}({\bf x},t)  \rangle}_T$, obtained numerically from a linear approximation based on time averaged velocities: $\lambda \nabla^2 {\bf b} = - {\bf B_0} ~ \nabla {\langle {\bf v}({\bf x},t) \rangle}_T $. This is similar to numerical  studies based on the averaged flow geometries~\cite{marie,bourgoinPOF}.  When ${\bf B}_0$ is applied along ${\bf \hat{z}}$, in a direction parallel to the rotation axis of the TG eddies, 
the most intense magnetic energy structures are concentrated round the $z=\pi/2,3\pi/2$ planes, in agreement with the differential rotation of the TG vortex. Moreover, the  most intense structures of ${\langle E_{b}({\bf x},t) \rangle}_T$ and ${\langle E_{b,lin}({\bf x},t) \rangle}_T$ fields coincide. For ${\bf B_0}$ along the ${\bf {\hat{x}}}$-axis, one observes  the main induction concentration around the $z=0, \pi$ planes, as expected from a direct inspection from the flow forcing. However, the most intense structures of the ${\langle E_{b}({\bf x},t) \rangle}_T$ and  ${\langle E_{b,lin}({\bf x},t) \rangle}_T$ fields do not coincide everywhere in that case (see location $(\pi/2,\pi/2,0)$ in Fig.~2(bottom), for example). Note also that the linear calculation over-estimates the time averaged magnetic fluctuations, whatever the orientation of the applied field.  Altogether it shows than one should be cautious when using average velocity fields in the calculation of magnetic induction, particularly if restricted to linear effects. The difference between the fields is probably linked to the large scale electromotive force due to turbulent motions. The influence of this force, as well as  the large scale induction topology  and its connection with the small scale fluctuations, will be reported in a forthcoming paper~\cite{prepa}.
\begin{figure}
\includegraphics[width=6cm]{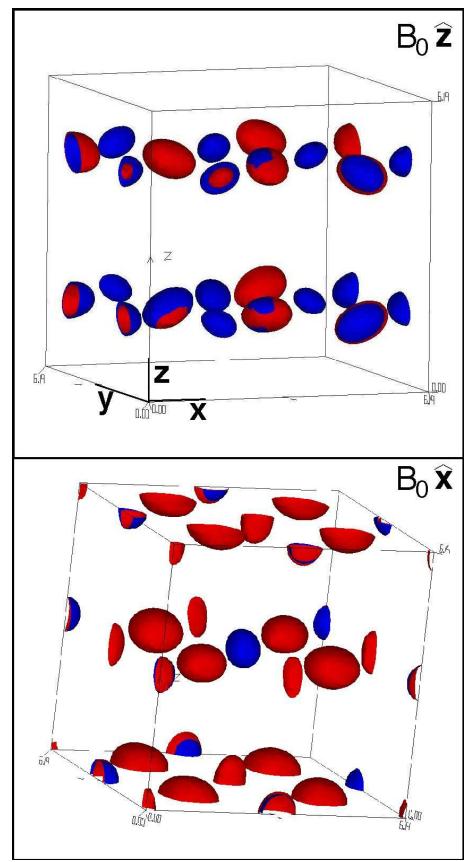}
\caption{Topology of the local induced magnetic energy, averaged in time, when ${\bf B}_0$ is applied along the ${\bf \hat{z}}$-axis (top) and along the ${\bf \hat{x}}$-axis (bottom) - in red: ${\langle E_{b}({\bf x},t) \rangle}_T$; in blue: ${\langle E_{b,lin}({\bf x},t) \rangle}_T$ - (see text). The isosurfaces are plotted at $80\%$ of the maximum values of the fields :
${\rm max} {\langle E_{b} \rangle}_T = 0.0056$  and  
${\rm max} {\langle E_{b,lin} \rangle}_T=0.0063$ for ${\bf B_0}=0.1~{\bf \hat{z}}$, 
and ${\rm max} {\langle E_{b} \rangle}_T =0.0041$ and 
${\rm max} {\langle E_{b,lin} \rangle}_T=0.0063$ for  ${\bf B_0}=0.1~{\bf \hat{x}}$.
}
\end{figure}

Figure~3 shows the temporal fluctuations of the induced field amplitude,  $|{\bf b}({\bf x},t)|$, probed inside the flow at two locations chosen from the previous topological observations, for ${\bf {B_0}}$ along the ${\bf \hat{x}}$-axis.  This is equivalent to using  local probes as in laboratory experiments.
\begin{figure}[b!] 
\includegraphics[totalheight=4cm]{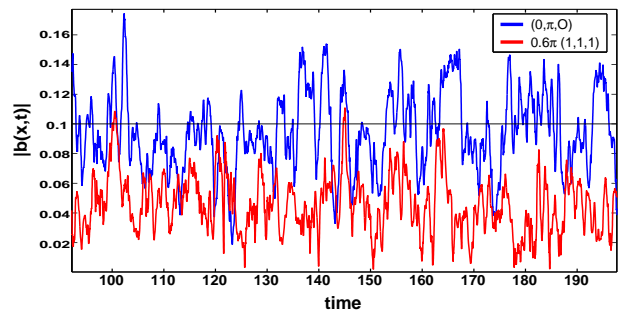} 
\caption{Time traces of $|{\bf b}({\bf x},t)|$, for  ${\bf B_0} = 0.1 \; {\bf \hat{x}}$,  at two  fixed points. In blue: $(0,\pi,0)$, mean value $\langle |{\bf b}({\bf x},t)| \rangle_T / B_0 = 0.92$, fluctuation level $|{\bf b}({\bf x},t)|_{rms} / B_0 = 0.28$. In red: $(0.6\pi, 0.6\pi, 0.6\pi)$ mean value $\langle |{\bf b}({\bf x},t)| \rangle_T / B_0 = 0.44$, fluctuation level $|{\bf b}({\bf x},t)|_{rms} / B_0 = 0.19$.}
\end{figure}
The intensity of the induced magnetic field has strong local fluctuations. The point at $(0, \pi, 0)$ is in a region of strong mean induction, whereas the point at $(0.6\pi, 0.6\pi, 0.6\pi)$ is at location of low mean induction (cf. Fig.~2(bottom)). We observe that, occasionally, the induced field gets larger than the applied field. In fact, if small amplitude fluctuations (about 10\%) are induced over time intervals of the order of the diffusive time $t_0$, much larger variations ($\sim 300$\%) can be observed over long time periods, of the order of $10 t_0$. These observations are in excellent qualitative agreement with the experimental observations at comparable $R_m$ and $P_m$~\cite{lathropspec,vkspof,karlspec,odier}. In order to be more quantitative, we analyze the time spectra; we focus on the case with ${\mathbf B}_0$ applied along the $\hat{\bf x}$-axis, but the results are identical when ${\mathbf B}_0$  is along $\hat{\bf z}$.  We plot in Figure~4 the power spectra of the temporal fluctuations of the magnetic field component $b_x({\bf x},t)$  recorded at $(0,\pi,0)$.  The higher end of the time spectrum follows a  behavior close to $f^{-11/3}$, as can be expected from the spatial spectrum using the Taylor  hypothesis of ``frozen''  field lines advected by the mean flow~\cite{odier}. In addition, for frequencies roughly between $1/t_0$  and $1/10 t_0$, 
the time spectrum develops a $1/f$ behavior, as observed in experimental measurements~\cite{vkspof}.  It is not present on the spatial spectrum in Figure~1, 
and thus appears as a distinctive feature of the {\it time dynamics} of the induced field. It is also independant of dynamo action, as it is also observed in the Karlsruhe experiments~\cite{karlspec}. 
\begin{figure}[ht] 
\includegraphics[height=3cm]{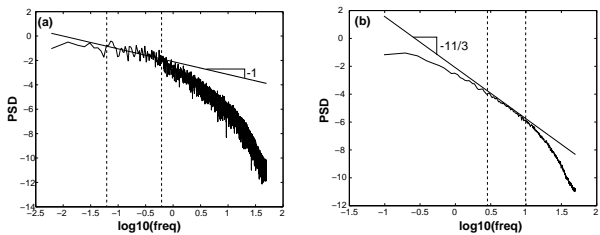} 
\caption{Power spectral density of the magnetic field fluctuations of $b_x({\bf x},t)$ 
in time, recorded at space location $(0,\pi,0)$, when ${\bf B_0} = 0.1 \; {\bf \hat{x}}$. 
(a) PSD computed as averages over Fourier transforms calculated over long time intervals ($\sim 164 t_0$) to emphasize the low frequency behavior; (b) PSD estimated from Fourier transforms over shorter time intervals ($\sim 10 t_0$). The behavior is identical for the $b_y({\bf x},t)$ and $b_x({\bf x},t)$ field components.} 
\end{figure}
Finally, our numerical study reveals one remarkable feature: the $1/f$ behavior is a global feature. It is observed on the fluctuations of the magnetic energy, as shown in Figure~5 (as a $f^{-2}$ scaling regime). We thus propose that it results from induction processes which have contributions up to the largest scale in the system. 
\begin{figure}[ht] 
\includegraphics[width=8cm]{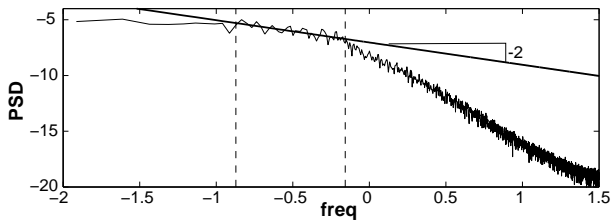} 
\caption{Power spectral density of the time fluctuations of the magnetic energy 
$E_b(t)=\langle {\bf b}^2(t) \rangle/2$, intergrated over space.} 
\end{figure}

To summarize, the mixed numerical scheme proposed here proves to be a valuable tool for the study of magnetohydrodynamics at low magnetic Prandtl numbers. We have considered here the response to an externally applied field.  The time behavior of magnetic field fluctuations is found in excellent agreement with experimental measurements. It has also revealed that the $1/f$ regime detected locally traces back to the global dynamics of the flow. Future work will analyze the contribution of turbulent fluctuations to the large scale magnetic field dynamics, and the influence of the magnetic Prandtl number on the threshold of the dynamo instability.\\

\noindent{\bf Acknowledgements:} We thank J.-P. Bertoglio, P. Odier and A. Pouquet for fruitful discussions. This work is supported by CNRS ATIP/SPI, PCMI and GdR-Dynamo. Computations performed on an Alineos PC cluster (OCA) and at IDRIS.

\end{document}